\documentclass[%
superscriptaddress,
nofootinbib,
amsmath,amssymb,
showkeys, twocolumn, prl]{revtex4-1}
\usepackage[dvipsnames]{xcolor}
\usepackage{amsmath,amsfonts,amssymb,amsthm, tabu}
\usepackage{xspace}
\usepackage{hyperref}
\usepackage{dsfont}
\usepackage{tikz}
\usetikzlibrary{arrows, decorations.markings, decorations.pathreplacing}
\usepackage{mathrsfs}
\usepackage{lipsum}
\usepackage{braket}
\usepackage{enumitem}
\usepackage{pgfplots}
\usepackage{pgf}
\pgfplotsset{compat=1.11}
\usepgfplotslibrary{fillbetween}
\usetikzlibrary{patterns}
\usepackage{amsbsy}
\usepackage{caption}
\usepackage{subcaption}
\usepackage[title]{appendix}
\usepackage{ifthen}
\usepackage{diagbox}
\usepackage{multibib}

\newcites{SM}{SM References}

\newcommand{\R}{\ensuremath{\mathbb{R}}}

\newcommand{\avgoe}[1]{\ensuremath{\left\langle #1\right\rangle}\xspace}
\newcommand{\tr}[1]{\ensuremath{\textrm{Tr}\left(#1\right)}}
\newcommand\Matching[4]{%
    \foreach \x/\y/\z/\w in {#3} {
    {\ifthenelse{\z = \w}
        {\ifthenelse{\z = 0}
        {\draw(-\y+1,#4) to[bend left=60] (-\x+1,#4) ;}
        {\draw(\x,#4) to[bend left=60] (\y,#4);}}
        {\ifthenelse{\z = 0}
        {\draw(-\x+1,#4) to[bend left=60] (\y,#4) ;}
        {}}
    }
    }
    \foreach \x in {1,...,#1}{
       \draw[circle, opacity=1, fill=white] (-\x+1,#4)circle[radius=0.75mm]node[below]{${\x}$};
    }
    \foreach \x in {1,...,#2}{
       \draw[circle,fill] (\x,#4)circle[radius=0.75mm]node[below]{${\x}$};
    }
}
\newcommand\ddfrac[2]{{\displaystyle\frac{\displaystyle #1}{\displaystyle #2}}}

\begin{document}

\title{Will Random Cone-wise Linear Systems Be Stable?}

\author{Th\'{e}o Dessertaine}
\affiliation{LadHyX UMR CNRS 7646, Ecole polytechnique, 91128 Palaiseau Cedex, France}
\affiliation{Chair of Econophysics \& Complex Systems, Ecole polytechnique, 91128 Palaiseau Cedex, France}

\author{Jean-Philippe Bouchaud}%
\affiliation{Chair of Econophysics \& Complex Systems, Ecole polytechnique, 91128 Palaiseau Cedex, France}
\affiliation{Capital Fund Management, 23 Rue de l'Universit\'{e}, 75007 Paris, France\medskip}

\date{\today}
\begin{abstract}
We consider a simple model for multidimensional cone-wise linear dynamics around cusp-like equilibria. We assume that the local linear evolution is either $\mathbf{v}^\prime=\mathbb{A}\mathbf{v}$ or $\mathbb{B}\mathbf{v}$ (with $\mathbb{A}$, $\mathbb{B}$ independently drawn a rotationally invariant ensemble of $N \times N$ matrices) depending on the sign of the first component of $\mathbf{v}$. We establish strong connections with the random diffusion persistence problem. When $N \to \infty$, we find that the Lyapunov exponent is non self-averaging, i.e. one can observe apparent stability and apparent instability for the same system, depending on time and initial conditions. Finite $N$ effects are also discussed, and lead to cone trapping phenomena.  
\end{abstract}
\maketitle

The stability of generic equilibrium points is well known to be determined by the largest eigenvalue of the matrix describing the linearised dynamics of small perturbations. However, a large variety of systems exhibit non-linearities even for infinitesimal perturbations. Such a situation arises when the dynamics involves threshold effects or constraints. For example, operational amplifiers in electrical systems involve diodes with voltage thresholds, whereas in neuroscience, simple models of neurons involve a gain function with a threshold over which the considered neuron will fire. For electrical engineering, the question of controlability of switch systems is also crucial. It was shown that, even in the simplest case, numerical assessment of stability and controlability is NP hard \cite{BLONDEL1999479}, see also \cite{CAMLIBEL20081261} for a detailed study of a particular class of switch systems. 

Similar effects have recently been discussed in the context of out-of-equilibrium macroeconomic models \cite{gualdi2015tipping, Dessertaine2021}. Economic equilibrium enforces that markets clear, i.e. that firms' supply $y$ equals households' demand $c$. However, in a dynamical setting, one may be in a situation where demand is -- say -- larger than supply. In this case, realized consumption $c_r$ is limited by production, i.e. $c_r = \min(c,y)$. Generically, the resulting update rules for firms' production and households' demand will differ when under-supply leads to excess savings or when over-supply leads to inventories. Hence, even small perturbations away from market clearing will evolve differently in the two regions $c > y$ and $c < y$ (see \cite{Dessertaine2021} for a generalisation to $n > 1$ firms). 

This is a situation we call ``cone-wise linear'': depending on the {\it direction} of the perturbation away from equilibrium, the linear stability matrix will not be the same. 
Even in the simplest case of a planar dynamical system with two cones, the overall dynamics can be highly non trivial and may generate limit cycles for example. For higher dimensional systems such as considered in  \cite{Dessertaine2021} (in the context of large economies), it is hard to get an intuition on the possible behaviors generated by such cone-wise dynamics. The purpose of the present work is to propose a simplified Random Matrix Theory framework to understand some of the phenomenology of these systems in the large dimension limit, in the spirit of R. May's celebrated study \cite{may1972will}. We find that the answer to the question ``will the system be stable?'' is not straightforward, with the emergence of complex non self-averaging behaviour.

\begin{figure}
    \centering
    \includegraphics{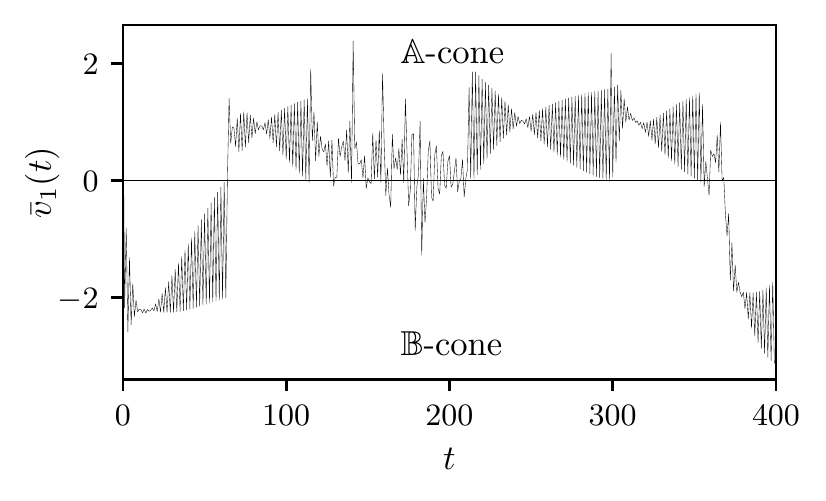}
    \caption{\small{Trajectory of the rescaled first component $\bar{v}_1(t)=\sqrt{N}v_1(t)/\|\mathbf{v}(t)\|\sim\phi(t)$ under the dynamics of Eq. \eqref{eq:cone-system} with $\mathbb{A}$ and $\mathbb{B}$ independently drawn GOE matrices with $N=10^3$, $\nu_{\pm,\mathbb{A}}=\nu_{\pm,\mathbb{B}}=\pm2$. We see that the dynamics is highly non-trivial with rapid oscillations caused by power of negative eigenvalues changing sign. Note also that the time spent in each cone varies widely from a single time-step to very long excursions.}}
    \label{fig:trajectory}
\end{figure}

The simple model we consider in this letter is the following: let $\mathbf{v}(t) \in\R^N$ be the $N$-dimensional vector describing the perturbation away from equilibrium. Depending on the sign of the dot-product $\mathbf{v}(t) \cdot \mathbf{e}$, where $\mathbf{e} \in\R^N$ is a {\it fixed} vector, the linearised dynamics is governed either by 
matrix $\mathbb{A}$ or by matrix $\mathbb{B}$, chosen to be symmetric $N \times N$ random matrices, independently drawn from  $O(N)$ rotationally invariant ensembles with possibly different eigenvalue spectra.\footnote{For $\mathbb{X}$ a rotationally invariant random matrix and $\mathbb{O}$ a rotation matrix, $\mathbb{X}\overset{d}{=}\mathbb{O}\mathbb{X}\mathbb{O}^\top$, where $\overset{d}{=}$ denotes the equality in distribution.}. In the present case, the cone structure is simply the two half-spaces separated by a hyperplane. We thus consider the following evolution for $\mathbf{v}(t)$\footnote{The case ${v}_1(t)=0$ has a zero probability for generic choices of $\mathbb{A}$ and $\mathbb{B}$, but if it where to happen, one would choose $\mathbb{A}$ or $\mathbb{B}$ with equal probability.} 
\begin{equation}
    \mathbf{v}(t+1)=\begin{cases}
    \mathbb{A}\mathbf{v}(t),&\text{when ${v}_1(t)>0$}\\
    \mathbb{B}\mathbf{v}(t),&\text{when ${v}_1(t)<0$}
    \end{cases},
    \label{eq:cone-system}
\end{equation}
where we have set $\mathbf{e}=(1,0,\dots,0)$ without loss of generality using the rotational invariance of both $\mathbb{A}$ and $\mathbb{B}$. The stability analysis therefore relies on the large time properties of matrix products of the type $\mathbb{M}(t)=\mathbb{A}^{t-t_{k-1}}\mathbb{B}^{\tau_{k-1}}\cdots\mathbb{A}^{\tau_1}$, with $t_{k-1}:=\sum_{i=1}^{k-1}\tau_i$ and $t_{k-1} < t \leq t_{k}$, and where $\tau_i$ are persistence times, i.e. times during which the dynamics leaves $\mathbf{v}(t)$ within the same cone (here the half space). It turns out that in our problem, the probability $Q_0(\tau)$ to remain in the same cone for a time $\geq \tau$ decays {\it algebraically}, $\tau^{-\mu}$ where $\mu$ is called the persistence exponent (see \cite{BrayReviewPersistence} for a detailed review). We will see that whenever $\mu < 1$ (which corresponds to natural choices for matrices $\mathbb{A}$ and $\mathbb{B}$, see below), the maximal Lyapunov exponent of the problem, namely
\begin{equation}
    \lambda_{\max}=\underset{t\to\infty}{\lim}t^{-1} \ln\left(\|\mathbb{M}(t) \mathbf{v}(t=0)\|/\|\mathbf{v}(t=0)\|\right),
\end{equation}
remains a random quantity even in the large time limit, and does not converge to its ensemble average value. In some cases one can observe $\lambda_{\max} < 0$, seemingly indicating stability, while in others (or at later times) $\lambda_{\max} > 0$, suggesting instability. This situation departs from the standard Furstenberg-Kesten result \cite{FurstenbergKesten} for products of random matrices. In fact, quite non-trivial dynamics can be observed, even for such a simple system, see Fig. \ref{fig:trajectory}.

Let us consider the evolution of $\mathbf{v}$ within one cone, say $v_1(t)>0$. As long as $v_1(t)>0$, the evolution is linear and yields $\mathbf{v}(t)=\mathbb{A}^t\mathbf{v}(0)$. Assuming that $\mathbf{v}(0)$ is a Gaussian random vector, the sequence of vectors $\mathbf{v}(t)$ is a centered Gaussian process with covariance given, after suitable scaling and in the large $N$ limit, by (see Supplementary Material)
\begin{equation}
    \avgoe{v_{\ell}(t)v_{\ell^\prime}(s)}=\delta_{\ell\ell^\prime}f_\mathbb{A}(t+s),
    \label{eq:large_n_gp}
\end{equation}
with,
\begin{equation}
f_\mathbb{A}(t)\underset{t\to\infty}{\sim}K\Gamma(\alpha+1) \, \nu_{+}^{t+\alpha+1}t^{-\alpha-1},
    \label{eq:f_asymp}
\end{equation}
where $K$ and $\alpha$ relate to the shape of the density of eigenvalue $\rho(\nu)$ of $\mathbb{A}$ near its upper edge $\nu_+$, to wit: $\rho(\nu)\underset{\nu\to\nu_{+}}{\sim}K|\nu-\nu_{+}|^{\alpha}$. Of course, a similar result holds for the covariance of the sequence of vectors induced by matrix $\mathbb{B}$. In the following, we will assume that the singularity exponent $\alpha$ is the same for $\mathbb{A}$ and $\mathbb{B}$ (but see Supplementary Material for the general case). Note that the natural case where $\mathbb{A}$ and $\mathbb{B}$ are (shifted) GOE matrices correspond to $\alpha=1/2$.

As in \cite{IIA_1}, we introduce the rescaled process $\phi(t)=v_1(t)/\sqrt{\avgoe{ v_1(t)^2}}$, for which one obtains the following asymptotic form for the correlator\footnote{Note that this correlator is independent of the constant $K$ appearing in Eq. \eqref{eq:f_asymp}. Correspondingly, the persistence probability $Q_0(\tau)$ is also independent of $K$.}
\begin{equation} \label{eq:correlator}
    \avgoe{\phi(t)\phi(s)}\underset{t,s\to\infty}{\sim}\left(\frac{2\sqrt{ts}}{t+s}\right)^{\alpha+1}.
\end{equation}
Interestingly, this is {\it exactly} the correlator of a well studied problem, namely the random diffusion process with an effective dimension $d=2(\alpha+1)$ \cite{2DdiffSchehr, DiffPersistenceNewman}. Consider the simple diffusion equation $\partial\phi/\partial t=\Delta\phi$ for a scalar field $\phi$ on $\R^d$ with random initial conditions having zero mean and short ranged correlation $\langle\phi(\mathbf{x},0)\phi(\mathbf{x}',0)\rangle=\delta^d(\mathbf{x}-\mathbf{x}')$. The probability that $\phi(\mathbf{0},t)$ does not change sign between time $t=0$ and $t=\tau$, is found to decay asymptotically as $\tau^{-\theta(d)}$, with a dimension dependent persistence exponent \cite{BrayReviewPersistence, CriticalDimensionsNewman}.

\begin{figure}[b]
    \centering
    \includegraphics{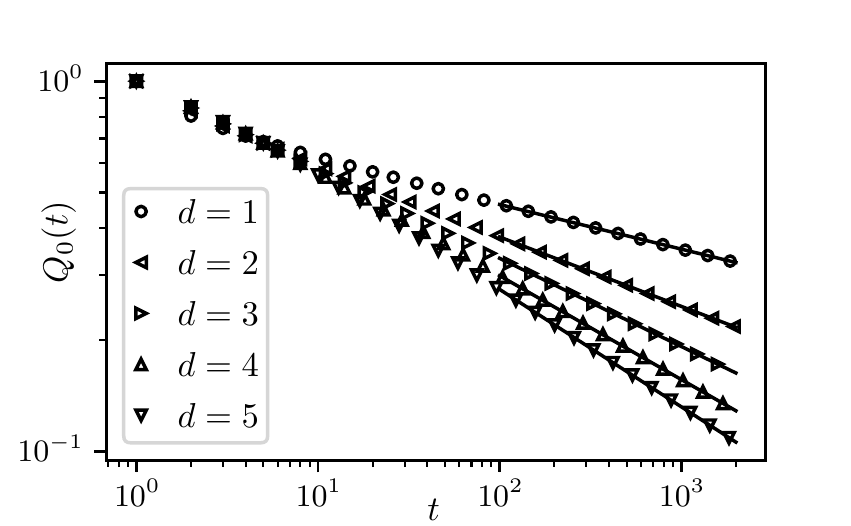}
    \caption{\small{Persistence probability of $\phi(t)$ for different effective dimensions $d$. The spectral distribution $\rho$ is chosen to be a standard symmetric Beta distribution $\text{{\fontfamily{lmss}\selectfont B}}\left(\frac{d}{2},\frac{d}{2}\right)$ with support $[0,1]$. The solid lines represent the algebraic decay of the diffusion process $\sim t^{-\theta(d)}$ using the persistence exponents computed in \cite{BrayReviewPersistence, 2DdiffSchehr}: $\theta(1)=0.1205$, $\theta(2)=3/16$, $\theta(3)=0.2382$, $\theta(4)=0.2806$, $\theta(5)=0.3173$. We averaged over the initial condition with $v_1(0)$ the first component of a normalized Gaussian vector.}}
    \label{fig:process_v_persistence_effective_dimension}
\end{figure}

 Now, whereas the Newell-Rosenblatt theorem \cite{NewellRosenblatt1962} allows us to surmise that there exists an exponent $\mu(\alpha)$ such that $Q_0(\tau)\sim \tau^{-\mu(\alpha)}$ for our vector persistence problem, the asymptotic equivalence of Eq. \eqref{eq:correlator} with the correlator of the random diffusion process does {\it not} necessarily imply equality of persistence exponents. However, in most cases, non-universal corrections to $Q_0(\tau)$, depending on the entire form of the correlator, are sub-leading in front of the algebraic decay inferred from the asymptotics and one can equate persistence exponents. Our numerical simulations strongly suggest that this is the case here, i.e. $\mu(\alpha)=\theta(d)$ with  $d=2(\alpha+1)$, see Fig. \ref{fig:process_v_persistence_effective_dimension}. In particular GOE matrices correspond to the random diffusion problem in $d=3$ dimensions. From the results of Ref. \cite{CriticalDimensionsNewman} on the $d$ dependence of $\theta$, we infer that for $\alpha\lesssim22$, the persistence exponent $\mu=\theta$ is less than unity, corresponding to an infinite mean survival time.

Now, let us come back to our product of random matrices problem $\mathbb{M}(t)=\mathbb{A}^{t-t_{k-1}}\mathbb{B}^{\tau_{k-1}}\cdots\mathbb{A}^{\tau_1}$, $t_k:=\sum_{i=1}^{k}\tau_i$. In the large $N$ limit, such products have been extensively studied in the context of free probabilities \cite{potters_bouchaud_2020, crisanti2012products, TucciProducts2011}. However, these methods do not apply here since the different terms in the product are not mutually free. In order to progress, we make the following independent interval hypothesis, namely that when the sign of $v_1(t)$ changes, i.e. at times $t_k$, the current vector $\mathbf{v}(t_k)$ can be considered as independent from the matrix ($\mathbb{A}$ or $\mathbb{B}$) under which it will evolve between $t_k$ and $t_{k+1}$. Therefore, persistence times $\tau_i$ can be considered as {\sc{iid}} random variables with distribution $\mathbb{P}(\tau_i=\tau):=p(\tau)=-\partial_\tau Q_0(\tau)$. (Again, we restrict here to the case where $\mathbb{A}$ or $\mathbb{B}$ share the same upper edge singularity exponent $\alpha$.) The second consequence of our independence assumption is that the growth of the norm of $\mathbf{v}(t)$ between $t=t_{k-1}$ and $t=t_{k}$ can be approximated, for large $N$, as 
\[
\|\mathbf{v}(t_{k})\|^2 =
{\sum_{a=1}^N w_a(t_{k-1}) \nu_a^{2\tau_k}} \approx 
\|\mathbf{v}(t_{k-1})\|^2 {\int {\rm d}\nu \rho_k(\nu) \nu^{2\tau_k}} 
\]
where $w_a = (\mathbf{v} \cdot \mathbf{u}_a)^2$ with $\mathbf{u}_a$ the eigenvectors of $\mathbb{A}$ or $\mathbb{B}$, depending on which of the two matrices is ``active'' between $t_{k-1}$ and $t_{k}$, and  $\rho_k(\nu)$ is the corresponding density of eigenvalues. 
We introduce the notation $g_k(\tau):=\frac12 \ln {\int {\rm d}\nu \rho_k(\nu) \nu^{2\tau}}$, such that the moments of the distribution of Lyapunov exponent $\lambda_{\max}$ can be expressed as
\begin{equation}
    \mathbb{E}\left[\lambda_{\max}^q\right]=\underset{t\to\infty}{\lim}t^{-q}\frac{\mathcal{Z}(q,t)}{\mathcal{Z}(0,t)},
    \label{eq:moments}
\end{equation}
where 
\begin{equation}
    \mathcal{Z}(q,t)=\sum_{k=1}^\infty\int_{\vec{\tau}} \prod_i p(\tau_i) \Theta\left(t_k-t\right)\Theta\left(t-t_{k-1}\right)\Lambda^q(\vec{\tau},t),
    \label{eq:partition_function}
\end{equation}
with $\vec{\tau}=(\tau_1,\ldots,\tau_k)$ and $
    \Lambda(\vec{\tau}, t):=\sum_{i=1}^{k-1}g_i(\tau_i)+g_k\left(t-t_{k-1}\right)$. Note that trivially $\mathcal{Z}(0,t)=1$.
    
in order to estimate the limit in \eqref{eq:moments}, we introduce the $t$-Laplace transform $\widehat{f}(\omega)=\int {\rm d}t \, e^{-\omega t}f(t)$ of a generic function $f$ and use Tauberian analysis to relate the large time behaviour of $f$ to the small $\omega$ behaviour of $\widehat{f}$. We first consider the case $\mu < 1$, in which case $p(t)\sim_{t\to\infty} Ct^{-1-\mu}$ implies $\widehat{p}(\omega)\sim_{\omega\to0}1+C\Gamma(-\mu)\omega^\mu$ (with $\Gamma(\cdot)$ the Euler's gamma function). Now, for \eqref{eq:moments} to have a finite non-trivial limit, one can make the following ansatz for $\mathcal{Z}(q,t)\underset{t \to \infty}{\sim} \mathbb{E}[\lambda_{\max}^q]t^{q}$. After finding a recursion relation for $\widehat{\mathcal{Z}}(q,\omega)$, one can show that the $q$-exponential generating function of $\widehat{\mathcal{Z}}$ denoted by $\mathcal{G}_{\widehat{\mathcal{Z}}}(x,\omega)$ verifies the equation
\begin{equation}
    \mathcal{G}_{\widehat{\mathcal{Z}}}=\frac{1}{2}\ddfrac{\mathcal{G}_{\widehat{h}_1}+\mathcal{G}_{\widehat{h}_2}+\mathcal{G}_{\widehat{h}_2}\mathcal{G}_{\widehat{\phi}_1}+\mathcal{G}_{\widehat{h}_1}\mathcal{G}_{\widehat{\phi}_2}}{1-\mathcal{G}_{\widehat{\phi}_1}\mathcal{G}_{\widehat{\phi}_2}},
\end{equation}
where $\widehat{\phi}_i(q,\omega)$ (resp. $\widehat{h}_i(q,\omega)$) is the Laplace transform of $t\mapsto p(t)g_i(t)^q$ (resp. $t\mapsto Q_0(t)g_i(t)^q$).


Using a scaling limit $\omega,x\to0$ while keeping the ratio $\omega/x=y$ constant, we can show that the probability density of $\lambda_{\max}$, denoted by $\varphi$, obeys the following equation
\begin{equation}
    \int d\lambda \, \frac{\varphi(\lambda)}{y-\lambda}=\frac{\left(y-r_1\right)^{\mu-1}+\left(y-r_2\right)^{\mu-1}}{\left(y-r_1\right)^{\mu}+\left(y-r_2\right)^{\mu}},
    \label{eq:density_eq}
\end{equation}
with $r_1=\ln{|\nu_{+,\mathbb{A}}|}$, $r_2=\ln{|\nu_{+,\mathbb{B}}|}$ and for $y\geq\max{\left(r_1, r_2\right)}$. Finally, using Stieltjes inversion formula, one finds the following density
\begin{equation}
    \varphi(\lambda)=\left|r_2-r_1\right|\frac{\sin\mu\pi}{\pi}\frac{(z_1z_2)^{\mu-1}}{z_1^{2\mu}+z_2^{2\mu}+2(z_1z_2)^{\mu}\cos(\mu\pi)},
    \label{eq:density_lyap}
\end{equation}
for $z_i=|\lambda-r_i|$. This distribution was first obtained by Lamperti \cite{lamperti1958occupation} and revisited by Godr\`eche \& Luck \cite{Godreche_Luck} in the context of occupation time of renewal stochastic processes. A moment of reflection allows to understand why this distribution appears in our problem as well: when $\mu < 1$, the mean persistence time diverges, which means that the longest persistence time $\tau_i$ observed in the interval $[0,t]$ is of order $t$ itself. Hence, Eq. \eqref{eq:partition_function} is dominated by long persistence times, for which $g_i(\tau_i) \approx r_i\tau_i$. In other words, our problem indeed boils down to an occupation time problem, at least within our independent interval hypothesis.  

\begin{figure}
    \centering
    \includegraphics{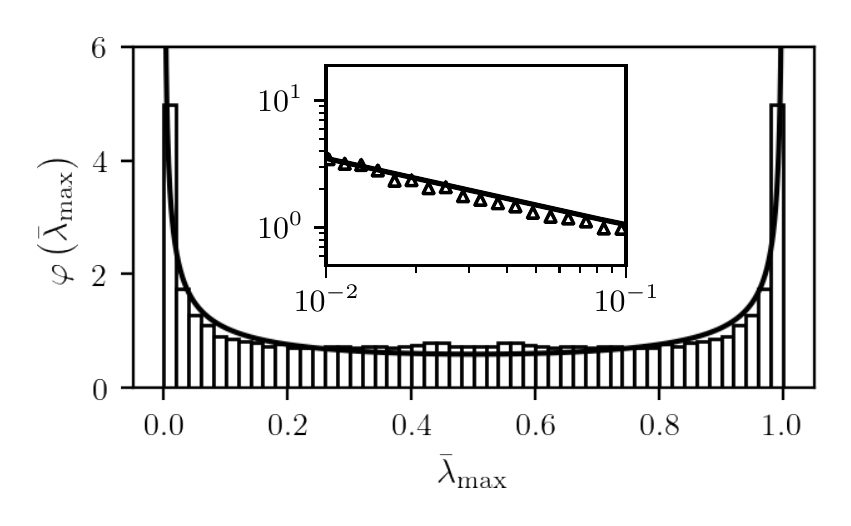}
    \caption{\small{Probability distribution of the normalized Lyapunov exponent $\bar{\lambda}_{\max}=(\lambda_{\max}-r_1)/(r_2-r_1)$ for GOE matrices $\mathbb{A}$ and $\mathbb{B}$ of size $N\times N$ with $N=10^{4}$ with $\nu_{+, \mathbb{A}}=0.05\sqrt{2}$ and $\nu_{+, \mathbb{B}}=2\sqrt{2}$. The black line shows the density Eq. \eqref{eq:density_lyap}. (Inset) Zoom on the left-tail of $\phi$ showing a very good agreement of the divergence with the predicted exponent $\mu-1$ (black line).}}
    \label{fig:Lyapunov_goe}
\end{figure}

Note that when $\lambda \to r_i$ the density $\varphi(\lambda)$ diverges as $K_i|\lambda-r_i|^{\mu-1}$, reflecting the dominance of long periods where the evolution is given either by matrix $\mathbb{A}$ (contributing to $\lambda \approx \ln|\nu_{+,\mathbb{A}}|$) or by matrix $\mathbb{B}$ (contributing to $\lambda \approx \ln|\nu_{+,\mathbb{B}}|$). Setting $\mu=1$ in Eq. \eqref{eq:density_eq}, we see that $\varphi(\lambda) \to \delta(\lambda-m_1)$ with $m_1=(r_1+r_2)/2$, rendering the system self-averaging. Fig. \ref{fig:Lyapunov_goe} shows the distribution of the Lyapunov exponent for matrices drawn from GOE and the density of Eq. \eqref{eq:density_lyap} with $\mu=2 \theta(d=3)$\footnote{Here, the persistence exponent doubles since Wigner's semi-circle distribution is symmetrical and therefore has odd moments equal to zero. As a consequence, $\left(v_\ell(2s)\right)_s$ and $\left(v_\ell(2s+1)\right)_s$ are mutually independent and the persistence of the entire system is the square of that of the even or odd process.}. The agreement with our theoretical prediction is very good. A rigorous hypothesis test is however difficult because of finite $N$ and $t$ effects that cannot be neglected, see below. 

In the case where $\mu > 1$ (i.e. $\alpha\gtrsim22$), the whole small $\omega$ analysis of $\widehat{\mathcal{Z}}(q,\omega)$ must be reconsidered and leads to the conclusion that the Lyapunov exponent $\lambda_{\max}$ becomes self-averaging and given by 
\begin{equation}
    \lambda_{\max}= \frac{\mathbb{E}\left[{g_{\mathbb{A}}(\tau)+ g_{\mathbb{B}}(\tau)}\right]}{2\mathbb{E}[\tau]},
\end{equation}
with $\tau$ distributed according to $p(\tau)$. However, observing this self-averaging regime is not straightforward since for $\alpha\gtrsim22$, the density of eigenvalues close to the upper edges is extremely small. 

\begin{figure*}[b]
    \centering
    \includegraphics{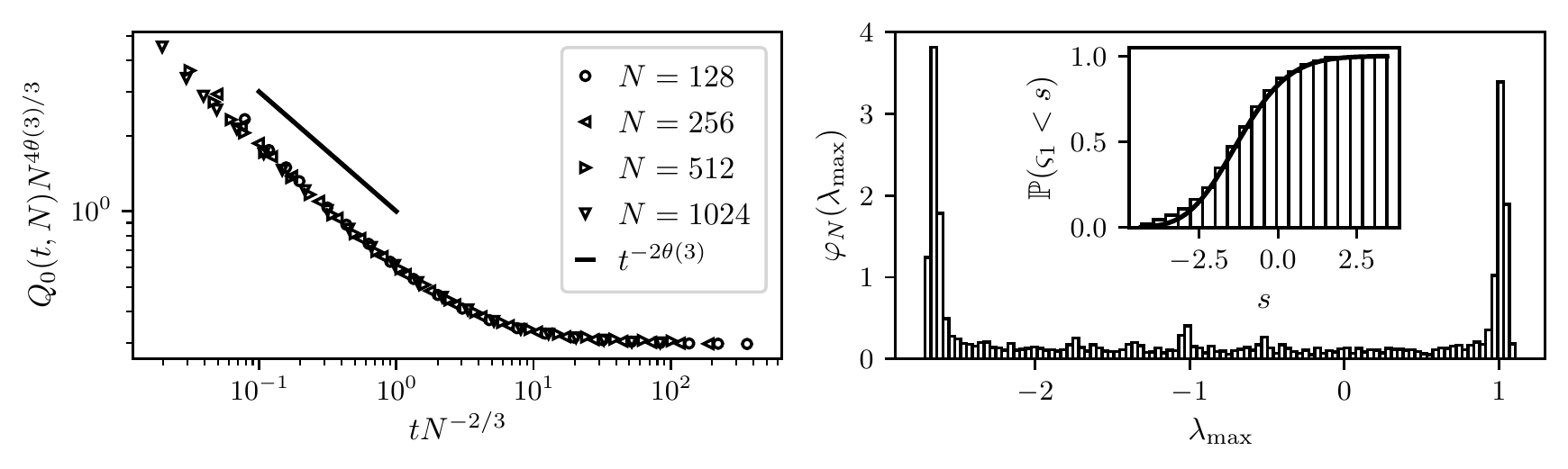}
    \caption{\small{(Left) Scaling form of the persistence probability at finite $N$ for matrices drawn GOE($N$). For GOE, $\alpha=1/2$ and $d=3$, however since the eigenvalue density is symmetrical, odd and even times processes uncorrelated, such that the persistence exponent is $2\theta(3)$. (Right) Probability distribution $\varphi_N$ of $\lambda_{\max}$ for finite $N=128$. We see that the edges of the spectrum are not sharp as in the case $N\to\infty$ but fluctuate around the corresponding values of $\ln{\nu_{+}}$ (with $\nu_{+, \mathbb{A}}=0.05\sqrt{2}$ and $\nu_{+, \mathbb{B}}=2\sqrt{2}$). (Inset) Cumulative distribution of the random variable $\varsigma_1=(e^{\lambda_{\max}}-\nu_{+})N^{2/3}/\gamma$ with $\gamma=\nu_{+}/2$. We overlay in black the cumulative distribution $F_1(s)$ of the $\beta=1$ Tracy-Widom distribution.} }
    \label{fig:scaling_finite_n_goe}
\end{figure*}

The previous analysis was conducted in the limit $N\to\infty$, before the large time limit $t \to \infty$ is taken. As we saw, in this limit, the persistence probability takes the form $Q_0(\tau)\sim \tau^{-\mu}$. However, whenever $N$ is finite, the eigenvalues of the matrices $\mathbb{A}$ and $\mathbb{B}$ do not perfectly sample the respective measures $\rho_{\mathbb{A}}(\nu)$ and $\rho_{\mathbb{B}}(\nu)$. Fluctuations near the edges of the spectrum are thus expected to change the persistence probability. As an example, let us consider matrices $\mathbb{A}$ drawn from the GOE. It is well known that at finite $N$, the maximum eigenvalue of $\mathbb{A}$ (or of $\mathbb{B}$) exhibits fluctuations of order $N^{-2/3}$ around the edge, abiding to the $\beta=1$ Tracy-Widom distribution $F_1$ \cite{Tracy:1992rf}. As a consequence,  following the analysis of \cite{2DdiffSchehr,MajumdarSchehr2007} for the random diffusion problem, we conjecture that the persistence probability at finite, large $N$ writes
\begin{equation}
    Q_0(\tau,N)\propto N^{-2\mu/3}h\left({\tau}{N^{-2/3}}\right),
    \label{eq:scaling_finite_N}
\end{equation}
where $h$ is a scaling function such that $h(u)\underset{u\to0}{\sim}u^{-\mu}$ and  $h(u)\underset{u\to\infty}{\rightarrow}c$, with $c$ a constant. Figure \ref{fig:scaling_finite_n_goe} (left) shows our numerical results that confirm such a scaling hypothesis, with again $\mu=2 \theta(3)$ for centred GOE matrices. The small $u$ behaviour of $h$ recovers the pure power law $\tau^{-\mu}$ in the large $N$ limit, whereas the large $u$ regime shows that, at finite $N$, $\mathbb{P}(\tau=\infty)= c >0$. This means that there is a positive probability that the vector $\mathbf{v}(t)$ remains ``trapped'' forever within a cone. 
But if the dynamics gets stuck within one cone, the associated Lyapunov exponent is simply given by $\lambda_{\max}=\ln{\nu_{\max,N}}$ where $\nu_{\max,N}$ is the largest eigenvalue of either $\mathbb{A}$ or $\mathbb{B}$ for a given finite value of $N$. As a consequence, the distribution of $e^{\lambda_{\max}}$ will have two peaks of width $N^{-2/3}$ centered around $\nu_{+, \mathbb{A}}$ and $\nu_{+, \mathbb{B}}$. The fluctuations around these peaks are given by Tracy-Widom distributions, as confirmed by our numerical data (see Figure \ref{fig:scaling_finite_n_goe} (right)).  

\onecolumngrid

\twocolumngrid

In conclusion, even our highly simplified cone-wise-linear dynamics exhibit quite non-trivial properties. First, if the large dimension limit $N \to \infty$ is taken before the large time limit, we find that for a wide class of random matrices the Lyapunov exponent is non self-averaging, i.e. continues to fluctuate in the large time limit. Depending on the relative positions of the upper edge of the spectrum of $\mathbb{A}$ and $\mathbb{B}$, one can observe apparent stability and apparent instability for the same system, depending on time and initial conditions. If, on the other hand, $N$ is large but finite, the dynamics eventually gets trapped in one of two half-spaces and the Lyapunov exponent converges to the top (log-)eigenvalue of the corresponding matrix $\mathbb{A}$ and $\mathbb{B}$. So if -- say -- $\mathbb{A}$ leads to a stable evolution and $\mathbb{B}$ to an unstable explosion, the dynamics of the system, even close to equilibrium, will be either stable or explosive, depending on minute details of the initial perturbation.

It would be of great interest to generalize our results in different directions. For example, for non symmetric Gaussian random matrices with a correlation coefficient $\varrho$ between entries $ij$ and $ji$ ($\varrho=1$ corresponding to the symmetric case), we have found numerically that the persistence probability $Q_0(\tau)$ decays as a truncated power-law $\tau^{-\mu(\varrho)}e^{-\tau/T(\varrho)}$ where $(\mu(\varrho),T(\varrho))$ interpolate between $(\mu,0)$ ($\varrho=1$: symmetric case) and $(0,1/\ln{2})$ ($\varrho=0$: Ginibre case). For generic $\varrho$, a non self-averaging behaviour occurs up to the time-scale $T(\varrho)$. More complicated cone structures could also be considered. The simplest generalisation is when the signs of $p$ vector components, say $v_1, v_2, \ldots, v_p$, select which of the $2^p$ matrices determine the dynamics. In this case, the persistence exponent is simply given by $p \mu$. Also, small $N$ effects should lead to a breakdown of our strong independence assumption, and generate even more complex types of dynamics, with non-random sequences of visited cones (see Supplementary Material). In view of the rich phenomenology reported in \cite{Dessertaine2021}, we believe that such effects would be well worth investigating.
\vskip .5cm

\begin{acknowledgments}
\small{\it{Acknowledgements}} \small{The authors wish to thank M. Benzaquen, J. Moran, J. Lumma and J. Garnier-Brun for their inputs. TD would also like to warmly thank P. Mergny for illuminating discussions on Random Matrix Theory. Finally, we are indebted to S. Majumdar and G. Schehr for their very interesting and useful comments. This research was conducted within the Econophysics \& Complex Systems Research Chair, under the aegis of the Fondation du Risque, the Fondation de l’Ecole polytechnique, the Ecole polytechnique and Capital Fund Management.}
\end{acknowledgments}


\bibliographystyle{unsrt}
\bibliography{biblio.bib}

\appendix

\pagebreak
\widetext
\begin{center}
\textbf{\large Supplementary Material: {\it Will Random cone-wise-Linear Systems Be Stable?}}
\end{center}
\setcounter{figure}{0}
\setcounter{page}{1}
\makeatletter
\renewcommand{\thefigure}{S\arabic{figure}}
\renewcommand{\bibnumfmt}[1]{[S#1]}
\renewcommand{\citenumfont}[1]{S#1}
\section{\texorpdfstring{Mapping to a Gaussian process and persistence of $v_1(t)$}{}}

In this section, we provide the details of the computation to map the dynamics of Eq. \eqref{eq:cone-system} of the main text onto a centered Gaussian process for which Eq. \eqref{eq:large_n_gp} is the particular case for symmetric matrices. Recall that we consider the following conewise-linear system
\begin{equation}
    \mathbf{v}(t+1)=\begin{cases}
    \mathbb{A}\mathbf{v}(t),&\text{when ${v}_1(t)>0$}\\
    \mathbb{B}\mathbf{v}(t),&\text{when ${v}_1(t)<0$}
    \end{cases},
\end{equation}
where $\mathbb{A}$ and $\mathbb{B}$ are $O(N)$ rotationally invariant random matrices. We assume that the initial condition $\mathbf{v}(0)\in\R^N$ is a centered Gaussian vector of unit variance. For the matrix $\mathbb{M}=\mathbb{A}\text{\;or\;}\mathbb{B}$, we denote by $\rho$ its spectral density and by $f(t)$ its $t$-th moment
\[f(t)=\int d\nu\rho(\nu)\nu^t.
\]
Finally, $\avgoe{(\;\cdot\;)}$ will denote the average over the \emph{disorder} $\mathbb{M}$, $\overline{(\;\cdot\;)}$ the average over the initial condition $\mathbf{v}(0)$ and $\tau(\;\cdot\;)=\lim_{N\to\infty}N^{-1}\tr{\cdot}$ the normalized trace.

\subsection{\texorpdfstring{Computation of the statistics of $\mathbf{v}(t)$ within one cone in the limit $N\to\infty$}{}}

As long as $v_1(t)$ keeps a constant sign, the dynamics is linear and immediately yields
\begin{equation}
    \mathbf{v}(t)=\mathbb{M}^t\mathbf{v}(0).
\end{equation}

Calling $\boldsymbol{\varepsilon_i}$ the $i$-th canonical vector of $\R^N$, we can express the $\ell$-th component $v_\ell(t)$ of $\mathbf{v}(t)$ as 
\begin{align*}
    v_\ell(t):=\boldsymbol{\varepsilon_\ell}\cdot\mathbf{v}(t)&=\sum_{\alpha}\left(\mathbb{M}^t\right)_{\ell\alpha}v_\alpha(0)
\end{align*}

Here, we consider a \emph{fixed} realization of the disorder $\mathbb{M}$. As a consequence, since $\mathbf{v}(0)$ is a Gaussian vector, one immediately sees that $\mathbf{v}(t)$ is also a Gaussian vector whose statistics can be computed easily. For the mean, we have
\begin{align*}
    \overline{v_\ell(t)}&=\sum_{\alpha}\left(\mathbb{M}^t\right)_{\ell\alpha}\overline{v_\alpha(0)}\\
    &=0
    \intertext{and for the covariance, we get}
    \overline{v_\ell(t)v_{\ell'}(s)}&=\sum_{\alpha,\beta}\left(\mathbb{M}^t\right)_{\ell\alpha}\left(\mathbb{M}^s\right)_{\ell'\beta} \overline{v_\alpha(0) v_\beta(0)}\\
    &=\sum_\alpha\left(\mathbb{M}^t\right)_{\ell\alpha}\left(\mathbb{M}^s\right)_{\ell'\alpha}\\
    &=\left(\mathbb{M}^t\left(\mathbb{M}^\top\right)^s\right)_{\ell\ell'}.
\end{align*}

Once again, these expressions are obtained for a fixed realization of the disorder $\mathbb{M}$. Taking the average over $\mathbb{M}$, one should usually be a bit careful. It is not straightforward that the average over $\mathbb{M}$ will leave the statistics of $\mathbf{v}(t)$ Gaussian. However in our case, we can see its correlator is actually self-averging in the limit $N\to\infty$ thanks to the rotational invariance of $\mathbb{M}$ and, as a consequence, the process remains Gaussian for $N\to\infty$ with covariance
\begin{equation}
    \overline{v_\ell(t)v_{\ell'}(s)}\underset{N\to\infty}{\sim}\frac{1}{N}\avgoe{\textrm{Tr}\,\left[\mathbb{M}^{t+s}\right]}\delta_{\ell\ell'}\underset{N\to\infty}{\longrightarrow}\tau\left(\mathbb{M}^t\left(\mathbb{M}^\top\right)^s\right)=\avgoe{\overline{v_\ell(t)v_{\ell'}(s)}}.
\end{equation}
We see that components are uncorrelated and, since they are Gaussian, therefore independent in the large $N$ limit. Finally, time-wise correlations are given by the \emph{mixed-moments} of matrices $\mathbb{M}$ and $\mathbb{M}^\top$.

\subsection{\texorpdfstring{Persistence of the sign of $v_1(t)$}{}}

From the previous section, we saw that, in the large $N$ limit, the components of $\mathbf{v}(t)$ become uncorrelated. Therefore, the probability that $\mathbf{v}(t)$ exits the cone after time $\tau$ is solely determine by the statistics of the first component and we have
\begin{equation}
    Q_0(\tau)=\mathbb{P}\left(\bigcap_{\tau=\tau+1}^\infty (v_1(0)v_1(\tau)>0)\right).
\end{equation}

\subsection{The symmetric case: link to random diffusion}

\begin{figure}
    \centering
    \includegraphics{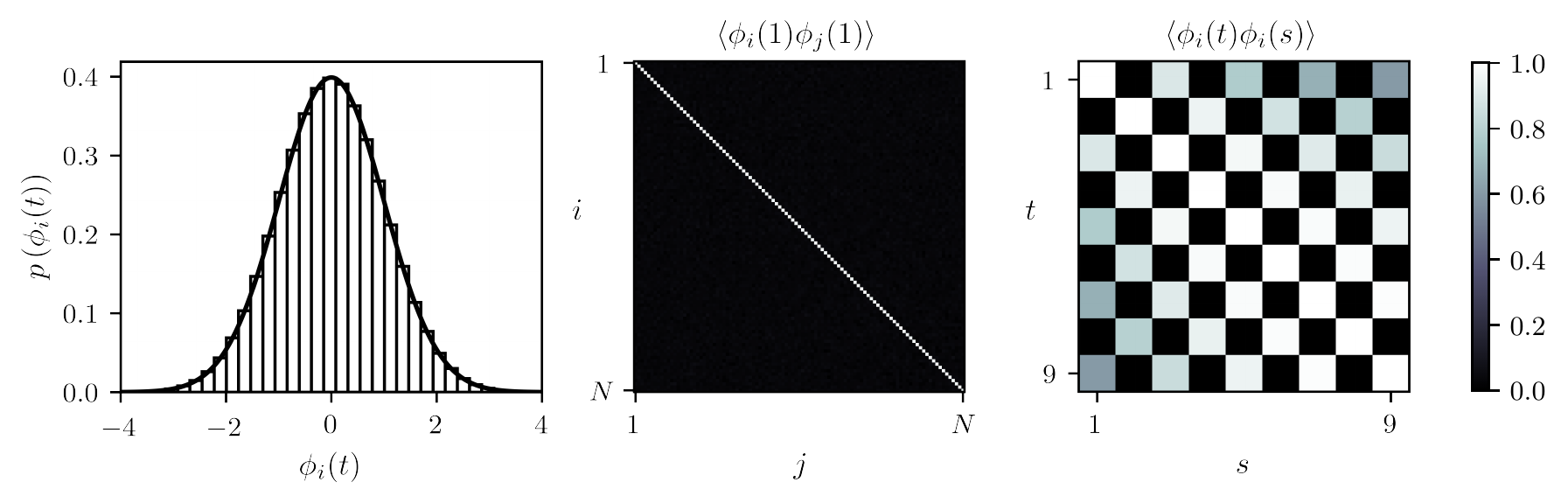}
    \caption{Simulations of the vector $\mathbf{v}(t)=\mathbb{M}^t\mathbf{v}(0)$ for $\mathbb{M}\in\textrm{GOE}(N)$, $\mathbf{v}(0)\hookrightarrow\text{{\fontfamily{lmss}\selectfont N}}\left(0,\mathbf{I}_N\right)$, $N=10^4$ and $t<10$. The average $\avgoe{(\;\cdot\;)}$ is performed over $5000$ realizations of $\mathbb{M}$ and $\mathbf{v}(0)$. (Left) Probability distribution of $\phi_i(t)$ overlaid with the density of a Gaussian random variable $\text{{\fontfamily{lmss}\selectfont N}}\left(0,1\right)$. (Middle) Component-wise covariance matrix $\langle\phi_i(t)\phi_j(t)\rangle$ at time $t=1$. We see that this covariance matrix is exactly $\mathbf{I}_N$ as predicted by Eq. \eqref{apeq:rescaled_gp}. (Right) Time-wise covariance matrix $\langle\phi_i(t)\phi_i(s)\rangle$ for $t,s<10$ which coincides with the prediction of Eq. \eqref{apeq:rescaled_gp}. Note the checkerboard structure coming from the fact that odd moments of GOE matrices are $0$.}
    \label{fig:processed_v}
\end{figure}

If the matrix $\mathbb{M}$ is symmetric, the mixed moment can be easily expressed
\begin{equation}
    \tau\left(\mathbb{M}^t\left(\mathbb{M}^\top\right)^s\right)=\tau\left(\mathbb{M}^{t+s}\right)=f(t+s),
\end{equation}
and amounts to the knowledge of the moments of $\mathbb{M}$. To establish the link with the random diffusion process, we must study the asymptotic behavior of the correlator of $\mathbf{v}(t)$, which boils down to the asymptotics of the moments of $\mathbb{M}$. In the previous section, we have introduced the spectral density $\rho$ of $\mathbb{M}$. As in the main text, we will assume that this density has a compact support $[\nu_-, \nu_+]$ with $\nu_-<\nu_+$. Furthermore, it has a behavior close to the upper edge charaterized by a constant $K$ and an exponent $\alpha$ which we assume greater than $-1$ for integrability purposes
\[\rho(\nu)\underset{\nu\to\nu_+}{\sim}K|\nu-\nu_+|^\alpha.\]
With these assumptions, we can use Laplace's method to estimate the large time behavior of the moments $f(t)$ of $\rho$. Denoting by $\Delta\nu=\nu_+-\nu_->0$, it yields
\begin{align}
    \nu_+^{-t}f(t)&=\int_{\nu_-}^{\nu_+}d\nu\rho(\nu)\left(\frac{\nu}{\nu_+}\right)^t\nonumber\\
    &=\int_0^{\Delta\nu}d\sigma\rho(\nu_+-\sigma)\left(1-\frac{\sigma}{\nu_+}\right)^t\nonumber\\
    &=\frac{\nu_+}{t}\int_0^{t\Delta\nu}dx\rho\left(\nu_+\left(1-\frac{x}{t}\right)\right)\left(1-\frac{x}{t}\right)^t\nonumber\\
    &\sim\frac{\nu_+}{t}\int_0^{\infty}dx K\left|\nu_+-\nu_+\left(1-\frac{x}{t}\right)\right|^\alpha e^{-x}\nonumber\\
    &=K\left(\frac{\nu_+}{t}\right)^{\alpha+1}\int_0^{\infty}dx e^{-x}x^\alpha\nonumber\\
    &=K\left(\frac{\nu_+}{t}\right)^{\alpha+1}\Gamma(\alpha+1),\nonumber\\
    \intertext{so that we get the asymptotic behavior of the main text}
    f(t)&\underset{t\to\infty}{\sim}K\Gamma(\alpha+1)\,\nu_{+}^{t+\alpha+1}t^{-\alpha-1}.\label{apeq:asymptotics_moments}
\end{align}
We then introduce (as it is standard in persistence problems), the rescaled process
\[\phi_i(t)=\frac{\bar{v}_i(t)}{\sqrt{\avgoe{\bar{v}_i^2(t)}}},\]
with $\phi(t):=\phi_1(t)$ as in the main text. Its correlator is given by
\begin{equation}
    \avgoe{\phi_i(t)\phi_j(s)}=\delta_{ij}\frac{f(t+s)}{\sqrt{f(2t)f(2s)}}.
    \label{apeq:rescaled_gp}
\end{equation}
Note that $\phi_i(t)$ is a Gaussian random variable with $0$ mean and unit variance. Using the asymptotic behavior of Eq. \eqref{apeq:asymptotics_moments}, we therefore get
\begin{equation}
    \avgoe{\phi(t)\phi(s)}\underset{t,s\to\infty}{\sim}\left(\frac{2\sqrt{ts}}{t+s}\right)^{\alpha+1},
    \label{apeq:correlator}
\end{equation}
and recover the result stated in the main text. Figure \ref{fig:processed_v} shows the statistics of the process $\phi_i(t)$, aligned with the previous theoretical predictions.

\section{\texorpdfstring{Distribution of $\lambda_{\max}$}{}}

In this appendix, we derive the equation for the moments $q$ exponential generating functions of $\widehat{\mathcal{Z}}$  and the density $\varphi$ of $\lambda_{\max}$. As stated in the text in Eq. \eqref{eq:moments}, the starting point is 
\begin{equation}
    \mathbb{E}\left[\lambda_{\max}^q\right]=\underset{t\to\infty}{\lim}t^{-q}\frac{\mathcal{Z}(q,t)}{\mathcal{Z}(0,t)},
    \label{apeq:moments}
\end{equation}
where 
\begin{equation}
    \mathcal{Z}(q,t)=\sum_{k=1}^\infty\int_{\vec{\tau}}\prod_{i=1}^k p_i(\tau_i)\Theta\left(t_k-t\right)\Theta\left(t-t_{k-1}\right)\Lambda^q(\vec{\tau},t):=\sum_{k=10}\mathcal{X}_k(q,t),
    \label{eq:gce_partition_function_ap}
\end{equation}
with $t_k=\sum_{i=1}^k\tau_i$ and
\begin{equation}
    \Lambda(\vec{\tau}, t)=\sum_{i=1}^{k-1}g_i(\tau_i)+g_k\left(t-t_{k-1}\right).
\end{equation}
The partition function $\mathcal{Z}(q,t)$ corresponds to a grand canonical ensemble partition function. We will denote by $\widehat{f}$ the $t$-Laplace transform of a generic function $f$
\begin{equation}
\widehat{f}(\omega)=\int_0^\infty dte^{-\omega t}f(t).
\end{equation}
We also introduce the following quantities
\begin{align}
    \widehat{\phi}_i(\ell,\omega)&=\int_0^\infty dt e^{-\omega t}p_i(t)g_i(t)^\ell,\\
    \widehat{h}_i(\ell,\omega)&=\int_0^\infty dtp_i(t)\int_0^t dse^{-\omega s}g_i(s)^\ell\\
    &=\int_0^\infty ds e^{-\omega s}g_i(s)^\ell Q_{0,i}(s),\\
    \widehat{h}_i(0,\omega)&:=\frac{1-\widehat{p}_i(\omega)}{\omega},
\end{align}
where $\widehat{\phi}_i(0,\omega):=\widehat{p}_i(\omega)$. Finally, note that, unlike in the main text, there is a subscript $i$ on the persistence probabilities. This is the general case where matrices $\mathbb{A}$ and $\mathbb{B}$ can have different edge exponents $\alpha_{\mathbb{A}}$
and $\alpha_{\mathbb{B}}$.

\subsection{Canonical ensemble computation}

We introduce the following canonical partition function which we will need for the computation of $\mathcal{Z}(q,t)$
\begin{equation}
    Z(q,t)=\sum_{k=1}^\infty\int_{\vec{\tau}}\prod_{i=1}^k p_i(\tau_i)\delta\left(t-t_k\right)\Lambda^q(\vec{\tau},t):=\sum_{k=1}^\infty x_k(q,t).
    \label{eq:ce_partition_function_ap}
\end{equation}
We will express the exponential generating function $G_{\widehat{Z}}(x)$ of $\widehat{Z}$. To do so, we must find a recursive relation on the Laplace transform of $Z$. Let us start by computing the quantities $\widehat{x}_k(q,\omega)$. We have for $x_k$ 
\begin{align}
    x_k(q,t)&=\int_{\vec{\tau}}\delta\left(t-t_k\right)\prod_{i=1}^k p_i(\tau_i)\left(\sum_{j=1}^{k}g_j(\tau_j)\right)^q,\nonumber
    \intertext{where we have replace the last term in the $q$-power $g_k\left(t-t_{k-1}\right)$ by $g_k(\tau_k)$ as enforced by the Dirac delta constraint. We can carry out the computation by expanding the $q$-power using the multinomial theorem}
    &=\int_{\vec{\tau}}\delta\left(t-t_k\right)\prod_{i=1}^k p_i(\tau_i)\sum_{\ell_1,\ldots,\ell_k}\delta\left(\sum_{i=1}^k\ell_i-q\right)\binom{q}{\boldsymbol{\ell}}\prod_{j=1}^q g_j(\tau_j)^\ell_j\nonumber\\
    &=\sum_{\ell_1,\ldots,\ell_k}\delta\left(\sum_{i=1}^k\ell_i-q\right)\binom{q}{\boldsymbol{\ell}}\int_{\vec{\tau}}\delta\left(t-t_k\right)\prod_{j=1}^k p_j(\tau_j)g_j(\tau_j)^{\ell_j}.\nonumber\\
    \intertext{We can then compute the Laplace transform of $x_k$}
    \widehat{x}_k(q,\omega)&=\int_0^{\infty} dt e^{-\omega t}\sum_{\ell_1,\ldots,\ell_k}\delta\left(\sum_{i=1}^k\ell_i-q\right)\binom{q}{\boldsymbol{\ell}}\int_{\vec{\tau}}\delta\left(t-t_k\right)\prod_{j=1}^k p_j(\tau_j)g_j(\tau_j)^{\ell_j}\nonumber\\
    &=\sum_{\ell_1,\ldots,\ell_k}\delta\left(\sum_{i=1}^k\ell_i-q\right)\binom{q}{\boldsymbol{\ell}}\int_{\vec{\tau}}\prod_{j=1}^k p_j(\tau_j)g_j(\tau_j)^{\ell_j}e^{-\omega \tau_j}\nonumber\\
    &=\sum_{\ell_1,\ldots,\ell_k}\delta\left(\sum_{i=1}^k\ell_i-q\right)\binom{q}{\boldsymbol{\ell}}\prod_{j=1}^k \widehat{\phi}_j(\ell_j,\omega).\nonumber
    \intertext{To carry out the computation, we need to distinguish between even and odd values of $k$. Indeed, provided $v_1(0)>0$, the effective matrix of the system will look like
\begin{align*}
    \mathbb{M}(2s)&=\mathbb{B}^{\tau_{2s}}\mathbb{A}^{\tau_{2s-1}}\cdots\mathbb{B}^{\tau_{2}}\mathbb{A}^{\tau_{1}}\\
     \mathbb{M}(2s+1)&=\mathbb{A}^{\tau_{2s+1}}\mathbb{B}^{\tau_{2s}}\cdots\mathbb{B}^{\tau_{2}}\mathbb{A}^{\tau_{1}},
\end{align*}
which in turn implies $g_{2s}=g_2$ and $g_{2s+1}=g_1$. As a consequence, we have}
    \widehat{x}_{2s}(q,\omega)&=\sum_{\ell_1,\ldots,\ell_{2s}}\delta\left(\sum_{i=1}^{2s}\ell_i-q\right)\binom{q}{\ell_1,\ldots,\ell_{2s}}\widehat{\phi}_{2}(\ell_{2s},\omega)\prod_{j=1}^{2s-1} \widehat{\phi}_j(\ell_j,\omega)\nonumber\\
    &=\sum_{\ell=0}^q\binom{q}{\ell}\widehat{\phi}_{2}(\ell,\omega)\sum_{\ell_1,\ldots,\ell_{2s-1}}\delta\left(\sum_{i=1}^{2s-1}\ell_i-(q-\ell)\right)\binom{q-\ell}{\ell_1,\ldots,\ell_{2s-1}}\prod_{j=1}^{2s-1} \widehat{\phi}_j(\ell_j,\omega)\nonumber,\\
    \intertext{where the sum over $\ell_1,\ldots,\ell_{2s-1}$ is exactly $\widehat{x}_{2s-1}(q-\ell,\omega)$. A similar fact holds for $\widehat{x}_{2s+1}(q-\ell,\omega)$ and we get}
    \widehat{x}_{2s}(q,\omega)&=\sum_{\ell=0}^q\binom{q}{\ell}\widehat{\phi}_{2}(\ell,\omega)
    \widehat{x}_{2s-1}(q-\ell,\omega),\quad s\geq1\label{eq:canonical_even_x}\\
    \widehat{x}_{2s+1}(q,\omega)&=\sum_{\ell=0}^q\binom{q}{\ell}\widehat{\phi}_{1}(\ell,\omega) \widehat{x}_{2s}(q-\ell,\omega),\quad s\geq1\label{eq:canonical_odd_x}.
\end{align}

These two equations do not account for $\widehat{x}_1(q,\omega)$ but it is easy to compute $\widehat{x}_1(q,\omega)=\widehat{\phi}_1(q,\omega)$. We now introduce odd and even partition functions
\begin{align}
    Z^{e}(q,t)&=\sum_{s=1}^\infty x_{2s}(q,t)\\
    Z^{o}(q,t)&=\sum_{s=0}^\infty x_{2s+1}(q,t)\\
    Z(q,t)&=Z^{e}(q,t)+Z^{o}(q,t),
\end{align}
and sum equations Eqs. \eqref{eq:canonical_even_x}-\eqref{eq:canonical_odd_x} to get the Laplace transforms of $Z^{e}$ and $Z^{o}$
\begin{align}
    \widehat{Z}^e(q,\omega)&=\sum_{\ell=0}^{q}\binom{q}{\ell}\widehat{\phi}_2(\ell,\omega)\widehat{Z}^o(q-\ell,\omega)\label{eq:equation_even_canonical_partition}\\
    \widehat{Z}^o(q,\omega)&=\widehat{\phi}_1(q,\omega)+\sum_{\ell=0}^{q}\binom{q}{\ell}\widehat{\phi}_1(\ell,\omega)\widehat{Z}^e(q-\ell,\omega).\label{eq:equation_odd_canonical_partition}
\end{align}

We recognize exponential convolution equations and we introduce the exponential generating function of a sequence $c$
\[\mathcal{G}_c(x)=\sum_{q=0}^\infty c(q)\frac{x^q}{q!},\]
to solve Eqs. \eqref{eq:equation_even_canonical_partition}-\eqref{eq:equation_odd_canonical_partition}. We get
\begin{align*}
    \mathcal{G}_{\widehat{Z}^{e}}&=\mathcal{G}_{\widehat{\phi}_2}\mathcal{G}_{\widehat{Z}^{o}}\\
    \mathcal{G}_{\widehat{Z}^{o}}&=\mathcal{G}_{\widehat{\phi}_1}+\mathcal{G}_{\widehat{\phi}_1}\mathcal{G}_{\widehat{Z}^{e}}
\end{align*}
and finally
\begin{align}
    \mathcal{G}_{\widehat{Z}^{e}}&=\frac{\mathcal{G}_{\widehat{\phi}_1}\mathcal{G}_{\widehat{\phi}_2}}{1-\mathcal{G}_{\widehat{\phi}_1}\mathcal{G}_{\widehat{\phi}_2}}\label{eq:egf_even_canonical}\\
    \mathcal{G}_{\widehat{Z}^{o}}&=\frac{\mathcal{G}_{\widehat{\phi}_1}}{1-\mathcal{G}_{\widehat{\phi}_1}\mathcal{G}_{\widehat{\phi}_2}}.\label{eq:egf_odd_canonical}
\end{align}
From these two equations we could continue the computation and get a general recursive formula for $\widehat{Z}(q,\omega)$ but we only need Eqs. \eqref{eq:egf_even_canonical}-\eqref{eq:egf_odd_canonical} to express the Laplace transform of the grand canonical partition function. 

\subsection{Grand canonical ensemble computation}

The idea for the computation in the grand canonical ensemble is similar to the canonical computation. However, upon summing over $t$, the constraint $\Theta\left(t_k-t\right)\Theta\left(t-t_{k-1}\right)$ does not imply $t-t_{k-1}=\tau_k$ anymore but $t_{k-1}\leq t\leq t_k$ i.e $t=t_{k-1}+\tau,\,\tau\in[0,\tau_k]$. We can then compute the Laplace transform of $\mathcal{X}_k$

\begin{align*}
    \widehat{\mathcal{X}}_k(q,\omega)&=\int_0^{\infty} dt e^{-\omega t}\sum_{\ell_1,\ldots,\ell_k}\delta\left(\sum_{i=1}^k\ell_i-q\right)\binom{q}{\boldsymbol{\ell}}\int_{\vec{\tau}}\Theta\left(t_k-t\right)\Theta\left(t-t_{k-1}\right)\prod_{j=1}^k p_j(\tau_j)g_j(\tau_j)^{\ell_j}\\
    &=\sum_{\ell_1,\ldots,\ell_k}\delta\left(\sum_{i=1}^k\ell_i-q\right)\binom{q}{\boldsymbol{\ell}}\int_{\vec{\tau}}\prod_{j=1}^{k-1} p_j(\tau_j)g_j(\tau_j)^{\ell_j}\int_0^\infty dt e^{-\omega t}\Theta\left(t_k-t\right)\Theta\left(t-t_{k-1}\right)g_k\left(t-t_{k-1}\right)^{\ell_k}\\
     &=\sum_{\ell_1,\ldots,\ell_k}\delta\left(\sum_{i=1}^k\ell_i-q\right)\binom{q}{\boldsymbol{\ell}}\int_{\vec{\tau}}\prod_{j=1}^{k-1} p_j(\tau_j)g_j(\tau_j)^{\ell_j}e^{-\omega t_{k-1}}\int_0^{\tau_k} d\tau e^{-\omega \tau}g_k(\tau)^{\ell_k}\\
    &=\sum_{\ell_1,\ldots,\ell_k}\delta\left(\sum_{i=1}^k\ell_i-q\right)\binom{q}{\boldsymbol{\ell}}\widehat{h}_k(\ell_k,\omega)\prod_{j=1}^{k-1} \widehat{\phi}_j(\ell_j,\omega).
\end{align*}
As in the canonical computation, we need to distinguish between the parity of $k$ and we have 
\begin{align}
    \widehat{\mathcal{X}}_{2s}(q,\omega)&=\sum_{\ell=0}^q\binom{q}{\ell}\widehat{h}_{2}(\ell,\omega)
    \widehat{x}_{2s-1}(q-\ell,\omega),\quad s\geq1\label{eq:even_x}\\
    \widehat{\mathcal{X}}_{2s+1}(q,\omega)&=\sum_{\ell=0}^q\binom{q}{\ell}\widehat{h}_{1}(\ell,\omega) \widehat{x}_{2s}(q-\ell,\omega),\quad s\geq1\label{eq:odd_x}.
\end{align}
Note that on the r.h.s we have recognized the canonical $\widehat{x}$. For $k=1$, we also have $\widehat{\mathcal{X}}_{1}(q,\omega)=\widehat{h}_{1}(q,\omega)$. Introducing odd and even grand canonical partition functions $\mathcal{Z}^{e}$ and $\mathcal{Z}^{o}$, we have
\begin{align*}
    \mathcal{G}_{\widehat{\mathcal{Z}}^{e}}&=\mathcal{G}_{\widehat{h}_2}\mathcal{G}_{\widehat{Z}^{o}}\\
    \mathcal{G}_{\widehat{\mathcal{Z}}^{o}}&=\mathcal{G}_{\widehat{h}_1}+\mathcal{G}_{\widehat{h}_1}\mathcal{G}_{\widehat{Z}^{e}},
\end{align*}
and using Eqs. \eqref{eq:egf_even_canonical}-\eqref{eq:egf_odd_canonical} we get for $\mathcal{G}_{\widehat{\mathcal{Z}}}=\mathcal{G}_{\widehat{\mathcal{Z}}^{e}}+\mathcal{G}_{\widehat{\mathcal{Z}}^{o}}$
\begin{equation}
    \mathcal{G}_{\widehat{\mathcal{Z}}}=\frac{\mathcal{G}_{\widehat{h}_1}+\mathcal{G}_{\widehat{h}_2}\mathcal{G}_{\widehat{\phi}_1}}{1-\mathcal{G}_{\widehat{\phi}_1}\mathcal{G}_{\widehat{\phi}_2}}.
    \label{eq:efg_grand_canonical}
\end{equation}

Note that, as in Eqs. \eqref{eq:egf_even_canonical}-\eqref{eq:egf_odd_canonical}, Eqs. \eqref{eq:efg_grand_canonical} are not symmetric upon the interchange $1\leftrightarrow2$. This is only due to the fact that we have considered products starting with the matrix $\mathbb{A}$ in the previous analysis. However, these products starts with either $\mathbb{A}$ or $\mathbb{B}$ with probability $1/2$ given the initial draw of $v_1(0)$. As a consequence, one should symmetrize Eqs. \eqref{eq:efg_grand_canonical} to get 
\begin{equation}
    \mathcal{G}_{\widehat{\mathcal{Z}}}=\frac{1}{2}\ddfrac{\mathcal{G}_{\widehat{h}_1}+\mathcal{G}_{\widehat{h}_2}+\mathcal{G}_{\widehat{h}_2}\mathcal{G}_{\widehat{\phi}_1}+\mathcal{G}_{\widehat{h}_1}\mathcal{G}_{\widehat{\phi}_2}}{1-\mathcal{G}_{\widehat{\phi}_1}\mathcal{G}_{\widehat{\phi}_1}},
\end{equation}
as in the main text. We will see however, that this does not matter for the long time behavior of the system.

\subsection{\texorpdfstring{Stieljes transform of $\lambda_{\max}$}{}}

Taking $x=0$ in Eq. \eqref{eq:efg_grand_canonical}, we easily get
\begin{equation}
    \widehat{\mathcal{Z}}(0,\omega)=\omega^{-1},
\end{equation}
which, after Laplace inversion yields
\begin{equation}
    \mathcal{Z}(0,t)=1.
\end{equation}
As a consequence, for the limit in Eq. \eqref{apeq:moments} to be well defined, one must have the following asymptotic behavior
\begin{equation}
    \mathcal{Z}(q,t)\underset{t\to\infty}{\sim}t^{q}\mathbb{E}\left[\lambda_{\max}^q\right],
\end{equation}
which in turns implies in Laplace space
\begin{equation}
    \widehat{\mathcal{Z}}(q,\omega)\underset{\omega\to0^+}{\sim} q!\omega^{-1-q}\mathbb{E}\left[\lambda_{\max}^q\right].
\end{equation}
Taking the limit $\omega\to0^+$ in the r.h.s of Eq. \eqref{eq:efg_grand_canonical} while keeping the ratio $\omega/x=y$ constant yields
\begin{align*}
    \mathcal{G}_{\widehat{Z}}(x,\omega)&\underset{\omega\to0^+}{\approx}\sum_{q=0}^\infty\frac{\left(\omega y^{-1}\right)^q}{q!}q!\omega^{-1-q}\mathbb{E}[\lambda_{\max}^q]\\
    &=\frac{y}{\omega}\mathbb{E}\left[\frac{1}{y-\lambda_{\max}}\right].
\end{align*}
The last expectation is the Stieljes transform of $\lambda_{\max}$ and we get
\begin{equation}
    \mathcal{G}_{\widehat{\mathcal{Z}}}(x,\omega)\underset{\omega\to0^+}{\approx}\frac{y}{\omega}\int d\lambda \frac{\varphi(\lambda)}{y-\lambda},
\end{equation}
with $\varphi$ the density of $\lambda_{\max}$

\subsection{\texorpdfstring{Distribution of $\lambda_{\max}$ in the case where $\alpha_\mathbb{A}=\alpha_\mathbb{B}$ and $\mathbb{E}[\tau]=+\infty$}{}}

In this case, the persistence exponents are the same $\mu(\alpha_\mathbb{A})=\mu(\alpha_\mathbb{B}):=\mu<1$ ensuring $\mathbb{E}[\tau]=+\infty$. Note however that the scale factor $C_i$ (such that $p_i(t)\sim C_i t^{-1-\mu}$) could still be different. However, since the correlator of $\phi$ in Eq. \eqref{apeq:correlator} is independent of the constant $K$ of the behavior of $\rho$ near the upper edge, we conclude that $C_\mathbb{A}=C_\mathbb{B}:=C$.\\

To use Eq. \eqref{eq:efg_grand_canonical} to find the Stieljes transform of $\lambda_{\max}$, we need to compute the behavior of the Laplace transforms as $\omega\to0$. We start by linking the long-time behavior of $p$ and $g_i$ to the small-$\omega$ behavior of the related Laplace transforms
\begin{equation}
    \left.\begin{aligned}
    p_i(t)&\underset{t\to\infty}{\sim}\frac{C}{t^{1+\mu}}\\
    g_i(t)&\underset{t\to\infty}{\sim}r_i t
    \end{aligned}\right\}\longleftrightarrow\left\{\begin{aligned}
    \widehat{p}_i(\omega)&\underset{\omega\to0}{\sim}1+C\Gamma(-\mu)\omega^\mu\\
    \widehat{\phi}_i(q,\omega)&\underset{\omega\to0}{\sim}r_i^q C\Gamma(q-\mu)\omega^{\mu-q},\quad q\geq1\\
    \widehat{h}_i(q,\omega)&\underset{\omega\to0}{\sim}r_i^q C\frac{\Gamma(1+q-\mu)}{\mu}\omega^{\mu-q-1},\quad q\geq1
    \end{aligned}\right.
    \label{eq:tauberian_table}
\end{equation}

Using the same scaling limit $\omega/x=y$ in the e.g.f of $\widehat{\phi}_i$ and $\widehat{h}_i$, we have
\begin{align*}
    \mathcal{G}_{\widehat{\phi}_i}(x,\omega)&\underset{\omega\to0^+}{\approx}1+C\Gamma(-\mu)\left(\frac{\omega}{y}\right)^{\mu}(y-r_i)^{\mu}\\
    \mathcal{G}_{\widehat{h}_i}(x,\omega)&\underset{\omega\to0^+}{\approx}C\Gamma(-\mu)\left(\frac{\omega}{y}\right)^{\mu-1}(y-r_i)^{\mu-1}.
\end{align*}
Plugging these expressions into Eq. \eqref{eq:efg_grand_canonical}, we get
\begin{equation}
\int d\lambda \frac{\varphi(\lambda)}{y-\lambda}=\frac{\left(y-r_1\right)^{\mu-1}+\left(y-r_2\right)^{\mu-1}}{\left(y-r_1\right)^{\mu}+\left(y-r_2\right)^{\mu}},
    \label{eq:ap_density_eq}
\end{equation}
which we can invert using Stieljes inversion formula
\begin{equation}
    \varphi(\lambda)=(r_2-r_1)\frac{\sin\mu\pi}{\pi}\frac{(z_1z_2)^{\mu-1}}{z_1^{2\mu}+z_2^{2\mu}+2(z_1z_2)^{\mu}\cos(\mu\pi)},
\end{equation}
with $z_i=\left|\lambda-r_i\right|$.

\subsection{\texorpdfstring{Distribution of $\lambda_{\max}$ in the case $\alpha_\mathbb{A}\neq\alpha_\mathbb{B}$ and $\mathbb{E}[\tau]=+\infty$}{}}

In this case, persistence exponents are different. We will assume w.l.o.g that $\mu_1<\mu_2$ with $\mu_1$ (resp. $\mu_2$) associated with $\mathbb{A}$ (resp. $\mathbb{B}$). Heuristically, we can surmise that the system will spend more time in the cone associated with the persistence having the smallest exponent. As a consequence, we expect $\lambda_{\max}$ to be self-averaging with value $\ln{|\nu_{+,\mathbb{A}}|}$.\\

The same small-$\omega$ analysis can be performed upon adding the relevant subscripts to scale factors and persistence exponents yielding the e.g.f
\begin{align*}
    \mathcal{G}_{\widehat{\phi}_i}(x,\omega)&\underset{\omega\to0^+}{\approx}1+C\Gamma(-\mu_i)\left(\frac{\omega}{y}\right)^{\mu_i}(y-r_i)^{\mu_i}\\
    \mathcal{G}_{\widehat{h}_i}(x,\omega)&\underset{\omega\to0^+}{\approx}C\Gamma(-\mu_i)\left(\frac{\omega}{y}\right)^{\mu_i-1}(y-r_i)^{\mu_i-1}.
\end{align*}
Plugging these into Eq. \eqref{eq:efg_grand_canonical} and keeping only the lowest orders $\omega^{\mu_1}$, we get
\begin{equation}
    \int d\lambda \frac{\varphi(\lambda)}{y-\lambda}=\frac{1}{y-r_1},
\end{equation}
which in turn implies that
\begin{equation}
    \varphi(\lambda)=\delta(\lambda-r_1).
\end{equation}

\subsection{\texorpdfstring{Self-averaging of $\lambda_{\max}$ for $\mathbb{E}[\tau]<+\infty$}{}}

We will consider the case $\alpha_\mathbb{A}=\alpha_\mathbb{B}$ (the case $\alpha_\mathbb{A}\neq\alpha_\mathbb{B}$ can be treated in the exact same way). If $\mathbb{E}[\tau]<\infty$ then $\mu>1$. Let us assume that $\mu\in]1,2[$ in order to carry fewer terms in the Tauberian analysis. The only important feature is that $\mu>1$ thus ensuring the existence of $\mathbb{E}[\tau]$.\\

Since the persistence has now a finite first moment, the small-$\omega$ behavior of the Laplace transforms are modified. We will denote by $m_\tau=\mathbb{E}[\tau]$ and $m_i=\mathbb{E}[g_i(\tau)]$ (which is finite since $g_i(\tau)\sim_{\tau\to\infty} r_i\tau$).
\begin{equation}
    \left.\begin{aligned}
    p(t)&\underset{t\to\infty}{\sim}\frac{C}{t^{1+\mu}}\\
    g_i(t)&\underset{t\to\infty}{\sim}r_i t
    \end{aligned}\right\}\longleftrightarrow\left\{\begin{aligned}
    \widehat{p}_i(\omega)&\underset{\omega\to0}{\sim}1-\omega m_\tau+C\Gamma(-\mu)\omega^\mu\\
    \widehat{\phi}_i(1,\omega)&\underset{\omega\to0}{\sim}m_i+r_i C\Gamma(1-\mu)\omega^{\mu-1},\quad q\geq1\\
    \widehat{\phi}_i(q,\omega)&\underset{\omega\to0}{\sim}r_i^q C\Gamma(q-\mu)\omega^{\mu-q},\quad q\geq2\\
    \widehat{h}_i(q,\omega)&\underset{\omega\to0}{\sim}r_i^q C\frac{\Gamma(1+q-\mu)}{\mu}\omega^{\mu-q-1},\quad q\geq1
    \end{aligned}\right.
    \label{eq:tauberian_table_sa}
\end{equation}
We get for the e.g.f
\begin{align*}
    \mathcal{G}_{\widehat{\phi}_i}(x,\omega)&\underset{\omega\to0^+}{\approx}1+m_i\frac{\omega}{y}-\omega m_\tau+C\Gamma(-\mu_i)\left(\frac{\omega}{y}\right)^{\mu_i}(y-r_i)^{\mu_i}\\
    \mathcal{G}_{\widehat{h}_i}(x,\omega)&\underset{\omega\to0^+}{\approx}m_\tau +C\Gamma(-\mu_i)\left(\frac{\omega}{y}\right)^{\mu_i-1}(y-r_i)^{\mu_i-1}.
\end{align*}
Plugging these into Eq. \eqref{eq:efg_grand_canonical} and keeping only the lowest orders, we get
\begin{equation}
    \int d\lambda \frac{\varphi(\lambda)}{y-\lambda}=\ddfrac{1}{y-\frac{m_1+m_2}{2m_\tau}},
\end{equation}
which in turn implies that
\begin{equation}
    \varphi(\lambda)=\delta\left(\lambda-\frac{m_1+m_2}{2m_\tau}\right).
\end{equation}

\section{Trajectories generated by the dynamics in Eq. \eqref{eq:cone-system}}

\begin{figure}[t]
    \centering
    \includegraphics{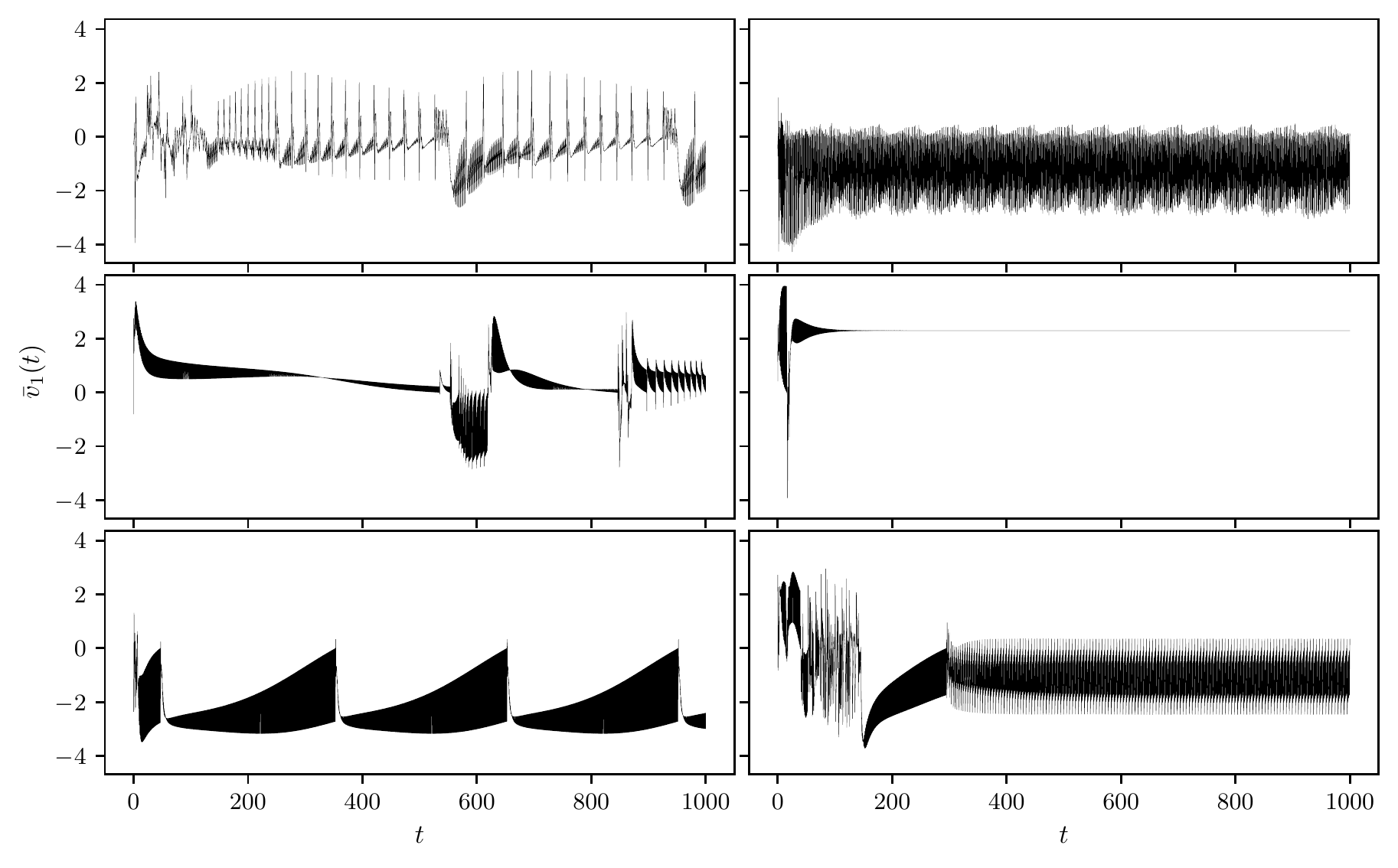}
    \caption{Different trajectories for $\bar{v}_1(t)=\sqrt{N}v_1(t)/\|\mathbf{v}(t)\|\sim\phi(t)$ where $\mathbf{v}(t)$ evolves according to Eq. \eqref{eq:cone-system} for $N=2500$ components. The initial condition $\mathbf{v}(0)$ is a Gaussian vector $\text{{\fontfamily{lmss}\selectfont N}}\left(0, \mathbf{I}_N\right)$. Matrices $\mathbb{A}$ and $\mathbb{B}$ are drawn from the Gaussian Orthogonal ensemble with $\nu_{\pm,\mathbb{A}}=0.05\sqrt{2}$ and $\nu_{\pm,\mathbb{A}}=2\sqrt{2}$.}
    \label{fig:other-dynamics}
\end{figure}

On Fig. \ref{fig:other-dynamics}, we give other examples of the type of complex dynamics that arise from the dynamical equation Eq. \eqref{eq:cone-system}. The simulations are performed for matrices $\mathbb{A}$ and $\mathbb{B}$ drawn from $GOE(N)$ with $N=2500$. We chose $\mathbb{A}$ to be a contraction matrix and $\mathbb{B}$ to be expanding. As we can see, the dynamical types observe are quite diverse. 

First-of-all, the dynamics gets stuck in the $\mathbb{A}$-cone on the middle-right plot. Indeed, since $N$ is finite, the persistence probability $Q_0(\tau)$ reaches a finite limit as $\tau\to\infty$ which indicates a non-zero probability of absorption. Not that here the absorption happens to take place in the contracting cone rendering the system stable. 

Second-of-all, on the bottom-left and top/bottom-right plots, the dynamics seems to reach limits cycles. As a consequence, the IIA approximation leading to the distribution of $\lambda_{\max}$ is not valid anymore since a "deterministic" sequence of cones is visited over and over. In this situation, $\lambda_{\max}$ should be equal (or at least very close to zero) and one should account for these dynamical type with an additional Dirac delta at zero in the expression of $\varphi$.

Finally, we can observe more complex dynamical types on the top/middle-left plots which are reminiscent of the dynamics presented on Fig. \ref{fig:processed_v} of the main text. 

\pagebreak

\bibliographystyleSM{unsrt}
\bibliographySM{biblio.bib}

\end{document}